\documentclass[12pt]{article}
\usepackage{graphicx}
\usepackage{subfigure}
\usepackage{latexsym}

\usepackage{color}

\newcommand{\beqs}{\begin{equation*}}

\newcommand{\eeqs}{\end{equation*}}

\newcommand{\beqas}{\begin{eqnarray*}}
\newcommand{\beqa}{\begin{eqnarray}}

\newcommand{\eeqas}{\end{eqnarray*}}
\newcommand{\eeqa}{\end{eqnarray}}

\newcommand{\blist}{\begin{itemize}}

\newcommand{\elist}{\end{itemize}}

\providecommand{\href}[2]{#2}

\DeclareFontFamily{OT1}{rsfs}{}
\DeclareFontShape{OT1}{rsfs}{m}{n}{ <-7> rsfs5 <7-10> rsfs7 <10->rsfs10}{} 
\DeclareMathAlphabet{\mycal}{OT1}{rsfs}{m}{n}

\def\TrL2{{\rm Tr}_{L^2}}

\def\atbdry{\Big|_{\partial \cM}}
\def\atbdry0{\Big|_{\partial \cM_0}}
\def\atbdry1{\Big|_{\partial \cM_1}}

\newcommand{\gs}{\mbox{$g_s$}}            
\newcommand{\ap}{\mbox{$\alpha^\prime$}}  
\newcommand{\ls}{\mbox{$l_s$}}            

\def\p{\partial}

\newcommand{\beq}{\begin{equation}}
\newcommand{\eeq}{\end{equation}}
\newcommand{\bea}{\begin{eqnarray}}
\newcommand{\eea}{\end{eqnarray}}

\begin{document}
\begin{titlepage}
\begin{flushleft}
       \hfill                     \\
       \hfill                       FIT HE - 13-01 \\
       \hfill                       
\end{flushleft}
\vspace*{3mm}
\begin{center}

{\bf\LARGE  Ad$S_5$ with two boundaries \\
 \vspace*{5mm}
  and holography of $\cal{N}=$4 SYM theory}


\vspace*{5mm}
\vspace*{12mm}
{\large Kazuo Ghoroku\footnote[2]{\tt gouroku@dontaku.fit.ac.jp} and 
Masafumi Ishihara\footnote[3]{\tt masafumi@wpi-aimr.tohoku.ac.jp},
Akihiro Nakamura\footnote[4]{\tt nakamura@sci.kagoshima-u.ac.jp}
}\\
\vspace*{2mm}

\vspace*{2mm}

\vspace*{4mm}
{\large ${}^{\dagger}$Fukuoka Institute of Technology, Wajiro, 
Higashi-ku}\\
{\large Fukuoka 811-0295, Japan\\}
\vspace*{4mm}
{\large ${}^{\ddagger}$WPI-Advanced Institute for Materials Research (WPI-AIMR), Tohoku University, Sendai 980-8577, Japan\\}

\vspace*{4mm}
{\large ${}^{\S}$Department of Physics, Kagoshima University, Korimoto1-21-35,Kagoshima 890-0065, Japan\\}

\vspace*{10mm}
\end{center}

\begin{abstract}

According to the AdS/CFT correspondence, 
the ${\cal N}=4$ supersymmetric Yang-Mills (SYM) theory is studied through its gravity dual whose
configuration has two boundaries at the opposite sides of the fifth coordinate. 
At these boundaries, in general, the four dimensional (4D) 
metrics are different, then we expect different properties for the theory living in two boundaries. 
It is studied how these two different properties of the theory 
are obtained from a common 5D bulk manifold in terms of the holographic method. 
We could show 
in our case that the two theories on the different boundaries are 
described by the Ad$S_5$, which is
separated into two regions by a domain wall.
This domain wall is
given by a special point of the fifth coordinate.
Some issues of the entanglement entropy related to this bulk configuration are 
also discussed. 
 
\end{abstract}
\end{titlepage}

\section{Introduction}

Up to now, many holographic approaches to the ${\cal N}=4$ supersymmetric 
Yang-Mills (SYM) theory have been performed in terms of the dual supergravity \cite{MGW}-\cite{CNP}. 
These approaches are based on a conjectured correspondence between a conformal field theory on the boundary 
M$_d$ of an asymptotic Anti de Sitter space {(AdS$_{d+1}$)} and string theories on the product of AdS$_{d+1}$ 
with a compact manifold. In many cases, the boundary M$_d$ is set as a Minkowski space-time,
and the bulk manifolds have a structure that they 
have a boundary at the ultraviolet (UV) side of the dual d-dimensional CFT. On the other hand, in the infrared side, 
they have a horizon. Then the holographic analyses for CFT in M$_d$
are performed in the region between the horizon and the boundary
for the gravity side. 

In these approaches, the research has been extended 
to the SYM theory in the background of $M_4=dS_4 (AdS_4)$ by 
introducing 4D cosmological constant
($\Lambda_4>0~ (\Lambda_4<0)$) \cite{H,GIN1,GIN2,EGR,EGR2,GN13}
in the supergravity solutions. In the bulk supergravity solutions, $\Lambda_4$
appears as a free parameter in the step of solving the equations of motion. However,
this parameter plays an important role since 
the 4D geometry of the boundary is controlled by this parameter. Another
important point is that the form of the bulk metric is also deformed by this parameter. 
As a result, we could see how the dynamical properties 
of the SYM theory are changed by the 4D geometry which is changed from the Minkowski
space-time to $dS_4 (AdS_4)$. 

Actually, in the cases of the dS$_4$ \cite{H,GIN1} and AdS$_4$ \cite{GIN2}, we 
find quite different properties of the SYM theory from the one observed in the 
Minkowsi space-time. For dS$_4$ background, we observe a horizon in the infrared
side of the fifth coordinate and 
we find the phenomena similar to the one of the finite temperature
SYN theory in the deconfinement phase. 
On the other hand, in the case of $\Lambda_4<0$  (for AdS$_4$ boundary), 
we could find that the theory is 
in the confining phase \cite{GIN2}. 
Further,  we found that the meson spectrum obtained in our analysis
is consistent with the one obtained in the usual field theory in AdS$_4$ \cite{AIS}.

Furthermore, we should notice that 
it is possible to introduce another free parameter in the bulk AdS$_5$ solution.
This parameter is called as the dark radiation, which corresponds to the thermal 
excitation of SYM fields, plays
an important role in determining the (de)confining phase of the theory \cite{EGR,EGR2,GN13}.

Our purpose in this article is to
point out and discuss a characteristic
holographic feature of the AdS$_5$ bulk solution. 
The point is that a second boundary appears in the solution with AdS$_4$ boundary
without horizons. 
It is found at
the "infrared limit" ($r=0$), 
\footnote{As shown below, the "infrared limit" doesn't mean the infrared limit of the SYM theory on the boundary.}
which is opposite to the boundary at "ultraviolet limit" ($r=\infty$). Here $r$ denotes the fifth coordinate of the AdS$_5$.
Then the gravity of the bulk AdS$_5$ is dual to the two theories living on the two boundaries
separately.

{We should notice that two boundaries are also seen in the case of the solution with 
dS$_4$ boundary. However, in this case,  
a horizon appears between the boundaries, then we can restrict
the region of the holographic dual to the one between the horizon and 
a boundary to study the dynamical
properties of the theory living on the boundary. In the other half region, the same things
are considered, however, the physics of the two boundaries might be
independent of each other.  \footnote{We could see similar situation in the case 
of the topological black hole solution in terms of the global coordinate $r$.}
In the case of AdS$_4$ boundary, on the other hand,
there is no horizon as such a border between the two boundaries 
}

\begin{figure}[htbp]
\vspace{.3cm}
\begin{center}
\includegraphics[width=14.0cm,height=7cm]{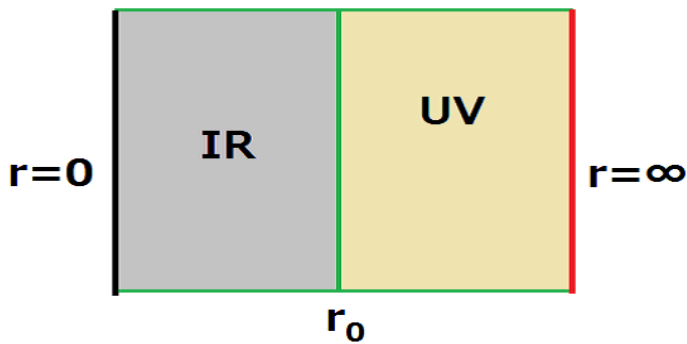}
\caption{A schematic picture which represents the bulk ${AdS_5}$
with two boundaries (shown by $r=0$ and $r=\infty$). {The middle line ($r=r_0$) shows
the domain wall (horizon) for  ${AdS_4}$ (${dS_4}$) boundary. This line
separates the bulk into two regions shown by "IR" and "UV", which are dual to the theory on
the boundaries at $r=0$ and $r=\infty $.}
\label{domain-wall}}
\end{center}
\end{figure}

The problem in this case {\bf (AdS$_4$ boundary)} is
how the bulk manifold could provide dynamical properties of the two field theories. In other words, 
how we could get information of two theories separately from the common
bulk geometry. This problem is resolved due to the presence of a sharp domain wall
in the bulk. The dynamical properties of the boundary theory are found through various stringy objects embedded in the bulk as probes
since the probes are controlled by the bulk configuration which reflects vacuum 
structure of the dual theory.
In the present case, we could find that the embedded objects are confined in one side and never
cross the wall to penetrate to the other side.
Then, in this sense, the gravity duals for the two boundaries are separated clearly by this wall. Therefore, this wall separates the
manifold to two regions which are surrounded by the boundaries at $r=0$ and $r=\infty$ respectively.
They correspond to two dual field theories of the two boundaries. The situation is shown
in the Fig.\ref{domain-wall}, where the two regions are shown  by "IR" and "UV".
We would address the holographic problem from this viewpoint and examine the robustness
of the wall.

This statement would be correct at the level of
classical in the gravity side of the bulk. 
When we consider
the quantum fluctuations of the bulk, they could cross the wall
since there is no obstacles to prevent their propagation like a singularity at this wall point.
On this problem, we will discuss in a future article.

\vspace{.3cm}
{When we add the dark radiation, our solutions of FRW type are modified. We find that
the role of the dark radiation is to shift the position of the horizon and the domain wall for
dS$_4$ and AdS$_4$ cases respectively. In the latter case, a phase transition 
from confinement to the deconfinement phase is seen when the 
magnitude of this term exceeds a critical value as shown in \cite{EGR}. 
Another observation is that
this term deforms the geometry of IR boundary.
On the other hand, the metric at the UV boundary  is not affected by the dark radiation.
This fact seems to be curious but interesting. We will give more details on this point in the future publication.  
}

\vspace{.5cm}
In the next section, our model to be examined is given and two boundaries of the gravity dual are shown.
Then, in the section 3, we show the existence of the domain wall which 
devides the bulk region of two boundary 
theories through the Wilson loop, D7 and D5 embeddings. From these, we can say quarks, flavored mesons 
and baryons 
are all separately examined in each bulk region corresponding to the dual theory in each boundary. 
In the section 4, the entanglement entropy is examined. In this case also, the minimal 
surface giving the entanglement entropy
of a theory in one boundary cannot penetrate into the region which is dual to the other boundary theory since
the penetration is protected by the domain wall. This fact implies that there is no entanglement of the
two theories of each boundary. 
Summary and discussions are given in the final section.

\section{Setup of the model}

First, we briefly review our model \cite{EGR,EGR2,GN13}.
We start from the 
10d type IIB supergravity retaining the dilaton
$\Phi$, axion $\chi$ and selfdual five form field strength $F_{(5)}$,
\beq\label{2Baction}
 S={1\over 2\kappa^2}\int d^{10}x\sqrt{-g}\left(R-
{1\over 2}(\partial \Phi)^2+{1\over 2}e^{2\Phi}(\partial \chi)^2
-{1\over 4\cdot 5!}F_{(5)}^2
\right), \label{10d-action}
\eeq
where other fields are neglected since {we do not need} them, and 
$\chi$ is Wick rotated \cite{GGP}.
Under the Freund-Rubin
ansatz for $F_{(5)}$, 
$F_{\mu_1\cdots\mu_5}=-\sqrt{\Lambda}/2~\epsilon_{\mu_1\cdots\mu_5}$ 
\cite{KS2,LT}, and for the 10d metric as $M_5\times S^5$,
$$ds^2_{10}=g_{MN}dx^Mdx^N+g_{ij}dx^idx^j=g_{MN}dx^Mdx^N+R^2d\Omega_{5}^2\, ,$$ 
we consider the solution. Here, the parameter is set as $(\mu=)1/R=\sqrt{\Lambda}/2$.

While the dilaton $\Phi$ and the axion $\chi$ play an important role when the bounadary
of $M_5$ is given by Minkowski space-time \cite{KS2,LT}, 
we neglect them here since we study the case
of (A)dS$_4$ boundary.
Then the equations of motion of non-compact five dimensional part
$M_5$ are written as
\footnote{The five dimensional $M_5$ part of the
solution is obtained by solving the following reduced 
Einstein frame 5d action,
\beq\label{action}
 S={1\over 2\kappa_5^2}\int d^5x\sqrt{-g}\left(R+3\Lambda
\right), \label{5d-action}
\eeq
which is written 
in the string frame and taking $\alpha'=g_s=1$ and the opposite sign
of the kinetic term of $\chi$ is due to the fact that
the Euclidean version is considered here \cite{GGP}.}

\beq\label{gravity}
 R_{MN}=-\Lambda g_{MN}\, .
\eeq
While this equation leads to the solution of Ad$S_5$, there are various Ad$S_5$ forms of the
solutions which are discriminated by the geometry of their 4D boundary as shown below.

\subsection{Solution}\label{sec22}

A class of solutions of the above equation (\ref{gravity})
are obtained in the following form of metric \cite{GN13}, 
\beq\label{10dmetric-2}
ds^2_{10}={r^2 \over R^2}\left(-\bar{n}^2dt^2+\bar{A}^2a_0^2(t)\gamma_{ij}(x)dx^idx^j\right)+
\frac{R^2}{r^2} dr^2 +R^2d\Omega_5^2 \ . 
\eeq
where
\beq\label{AdS4-30} 
    \gamma_{ij}(x)=\delta_{ij}\left( 1+k{\bar{r}^2\over 4\bar{r_0}^2} \right)^{-2}\, , \quad 
    \bar{r}^2=\sum_{i=1}^3 (x^i)^2\, ,
\eeq
and $k=\pm 1,$ or $0$. The arbitrary scale parameter  $\bar{r_0}$ is set hereafter as $\bar{r_0}=1$.
For the undetermined non-compact five dimensional part, the following equation is obtained from the
$tt$ and $rr$ components of (\ref{gravity}) \cite{BDEL,Lang},
\beq\label{A1}
 \left({\dot{a}_0\over a_0}\right)^2+{k\over a_0^2}=
   -{\Lambda\over 4}A^2+\left({{r\over R}A'}\right)^2
  +{C\over a_0^4 A^2}\ , 
\eeq
where $\dot{a_0}=\partial a_0/\partial t$, $A'=\partial A/\partial r$, and 
\beq
 A={r\over R}\bar{A} , \quad {\partial_t({a_0(t)A})\over \dot{a}_0(t)}={r\over R}\bar{n}\, .
\eeq
The constant $C$ is given as an integral constant in obtaining (\ref{A1}), 
and we could understand that it corresponds to
the thermal excitation of ${\cal N}=4$ SYM theory for $a_0(t)=1$, and  
{it is called as dark radiation \cite{BDEL,Lang}.  

At this stage, two undetermined functions, $\bar{A}(r,t)$ and $a_0(t)$, are remained. textcolor{red}{However} the equation to solve
them is the Eq.(\ref{A1}) only.  {Therefore, we could determine $a_0(t)$ 
by introducing the 4D Friedmann equation, 
which is independent of (\ref{gravity}). However it should be realized on the boundary where
various kinds of matter could be added in order to form the presumed FRW universe as in \cite{GN13}}
\bea\label{bc-RW2}
  \left({\dot{a}_0\over a_0}\right)^2+{k\over a_0^2}&=& 
 {\Lambda_4\over 3}+{\kappa_4^2\over 3}\left(
    {\rho_m\over a_0^3}+ {\rho_r\over a_0^4}+{\rho_u\over a_0^{3(1+u)}} \right)\equiv
        \lambda(t)\, \label{bc-RW3}
\eea
where $\kappa_4$ ($\Lambda_4$) denotes the 4D gravitational constant 
(cosmological constant).  The quantities $\rho_m$ 
and $\rho_r$ denote the energy density of 
the nonrelativistic matter and the radiation of 4D theory respectively. 
The {most right hand side} expression $\lambda(t)$ 
in (\ref{bc-RW3}) is given as a simple form of
the {most} left hand side of (\ref{bc-RW3}) given by 
using $a_0(t)$. Then the remaining
function $A(t,r)$ is obtained from (\ref{A1})
in terms of $\lambda(t)$. {The last term $\rho_u$ { in the middle of
(\ref{bc-RW3})} represents an unknown matter
with the equation of state, $p_u=u \rho_u$, where $p_u$ and 
$\rho_u$ denote {its} pressure and energy 
density respectively.}
It is important to be able to solve the bulk equation (\ref{A1}) in this way by relating
its left hand side to the Friedmann equation defined on the boundary \cite{GN13} since
we could have a clear image for the solution.

Finally, 
the solution 
is obtained as

\bea
 \bar{A}&=&\left(\left(1-{\lambda\over 4\mu^2}\left({R\over r}\right)^2\right)^2+\tilde{c}_0 \left({R\over r}\right)^{4}\right)^{1/2}\, , \label{sol-10} \\
\bar{n}&=&{\left(1-{\lambda\over 4\mu^2}\left({R\over r}\right)^2\right)
         \left(1-{\lambda+{a_0\over \dot{a}_0}\dot{\lambda} \over 4\mu^2}\left({R\over r}\right)^2\right)-\tilde{c}_0 \left({R\over r}\right)^{4}\over 
       \sqrt{\left(1-{\lambda\over 4\mu^2}\left({R\over r}\right)^2\right)^2+\tilde{c}_0 \left({R\over r}\right)^{4}}}\, , \label{sol-11}
\eea
where 
\beq
\tilde{c}_0=C/(4\mu^2a_0^4)\, . \label{sol-12}
\eeq


\subsection{Two boundaries}

\noindent{\bf Urtraviolet boundary $r\to\infty$}

In the case of the above solution, there is a boundary at $r\to\infty$, where
the energy scale of the dual field theory is at the ultraviolet limit. The boundary should
be set at the position where the metric has a second order pole \cite{KS} since 
the manifold is not well defined there.
At this boundary $r\to\infty$, the 4D metric is given as 
\beq\label{RW}
  ds_{\rm FRW}^2=-dt^2+a_0(t)^2\gamma_{ij}dx^idx^j\, ,
\eeq
since the above solution behaves as $\bar{n}\to 1$ and $\bar{A}(r,t)\to 1$ for $r\to\infty$. 
This is the well-known Friedmann-Robertson-Walker (FRW) metric, which is usually used in cosmology to study
the time development of our universe. In the present case, therefore, we can study the SYM theory
in this FRW universe from the bulk metric (\ref{10dmetric-2}) which is the holographic dual as 
shown in \cite{GN13}.

\vspace{.3cm} 
\noindent{\bf Infrared boundary $r\to 0$}

Next, from (\ref{sol-10}) and (\ref{sol-11}), we find that there is another boundary 
at $r\to 0$, in the infrared limit, for $\Lambda_4<0$, small $C$ and tiny time dependence of $\lambda(t)$.
However the appearance of this boundary depends on the time when the effect of $C$ and
time dependence of $\lambda(t)$ are considered. The situation is therefore a little complicated. 
For example, consider
the case of $\dot{\lambda}=0$ for simplicity, then the second boundary is found for $\lambda<0$ and 
$|{\lambda\over 4\mu^2}|>\sqrt{\tilde{c}_0} $. However, the last inequality depends on time and 
it is satisfied for restricted time-interval. So more simple case is considered below.

\vspace{.6cm} 
\noindent{\bf i) For the case of $C=0$ and negative constant $\lambda(=-\lambda_0)$}

In order to make clear the two boundaries, the situation is simplified 
by considering the case of $\lambda=-\lambda_0$ and
$C=0$, where $\lambda_0$ is a positive constant.  This is corresponding 
to the case of negative $\Lambda_4$ and $\lambda_0=-\Lambda_4/3$.  
In this case, the scale factor is given by solving Eq.(\ref{bc-RW2}) for $k=-1$ as follows,
\beq
    a_0(t)=\sin\left(\sqrt{\lambda_0}t\right)/\sqrt{\lambda_0}
\eeq
and then the metric is written as
\bea
ds^2_{10}&=&ds^2_{5}+R^2d\Omega_5^2 \label{AdS4-1} \\ 
   ds^2_{5}&=&{r^2 \over R^2}\left(1+{r_0^2\over r^2}\right)^2\left(-dt^2+a_0^2(t)\gamma_{ij}(x)dx^idx^j\right)+
\frac{R^2}{r^2} dr^2 \label{AdS4-2} 
\eea
where $r_0^2=\lambda_0R^4/4$ and
\beq\label{AdS4-3} 
    \gamma_{ij}(x)=\delta_{ij}\left( 1-{\bar{r}^2\over 4} \right)^{-2}\, .
\eeq
In this case, the boundary represents a typical AdS$_4$
manifold, and the SYM theory on this manifold has been holographically 
examined well previously \cite{GIN2}.
{In this case, the analysis has been performed by supposing that
the bulk is dual to the theory on the boundary $r=\infty$. And we have paid no attention
to the other possible boundary at $r=0$.}

However, {in order to have correct results of the analysis, we must
notice the fact that there is actually} another boundary at $r\to 0$ in the bulk of (\ref{AdS4-1}).
In order to see this point clearly, we rewrite the above metric by changing the coordinate $r$ as 
$r=r_0^2/z$, then we have
\beq\label{AdS4-4-In}
ds^2_{10}={z^2 \over R^2}\left(1+{r_0^2\over z^2}\right)^2
\left(-dt^2+a_0^2(t)\gamma_{ij}(x)dx^idx^j\right)+
\frac{R^2}{z^2} dz^2 +R^2d\Omega_5^2 \ . 
\eeq
Then we find again the same form of metric with (\ref{AdS4-1}), but $r$ is replaced by $z$. This implies 
the following two points. (i) There must be another bulk region near $z=\infty$
which is dual to SYM theory living on $AdS_4$ at $z=\infty$. (ii) Secondly, the limit of $z=\infty$ is also
the urtaviolet region of the SYM theory as understood from the form of (\ref{AdS4-4-In}).
Then we could obtain the same dynamical 
information, from the metric (\ref{AdS4-4-In}),
of the theory with the one given for the theory at $r=\infty$. 
In other words, in the bulk manifold, the same two
dual theories should be expressed by two regions which are separated at some point of the 
coordinate $r$. This point is called as domain wall, and we could find it at $r=r_0$
as shown below.  

\vspace{.6cm} 
\noindent{\bf ii) For the case of $C=0$ and positive constant $\lambda=\lambda_0$}

{In the case of $\lambda=\lambda_0>0$,  $\Lambda_4$ is positive and 
the scale factor is given by solving Eq.(\ref{bc-RW2}) for $k=0$ as follows,
\beq
    a_0(t)=a(0)e^{\sqrt{\lambda_0}t}\, ,
\eeq
and then the metric is written as
\bea
ds^2_{10}&=&d\tilde{s}^2_{5}+R^2d\Omega_5^2 \label{AdS4-1} \\ 
   d\tilde{s}^2_{5}&=&{r^2 \over R^2}\left(1-{r_0^2\over r^2}\right)^2\left(-dt^2+a_0^2(t)\delta_{ij}(x)dx^idx^j\right)+
\frac{R^2}{r^2} dr^2 \label{AdS4-2} \, .
\eea
In this case, the boundary represents dS$_4$
manifold, and the SYM theory on this manifold has been holographically 
examined well previously \cite{GIN1} for the
theory on the boundary $r=\infty$. And we have considred only for the half region
of $r_0<r<\infty$, then no attention is paid to the other possible boundary at $r=0$.
In this case, however, the situation is different from the above case, 
and it would be reasonable to restrict to the region $r_0<r<\infty$ 
since the point $r=r_0$ represent the horizon. The situation is similar to the case
of the Schwartzschild-AdS background, where the holographic region is restricted
to the region from the horizon to $r=\infty$.

It would be an interesting problem to study the theory ar $r=0$ boundary by considering
the region of $0<r<r_0$. We expect similar behaviour to the theory on $r=\infty$.
However, here, we give such study in the future article. 
}

\vspace{.3cm}
\noindent{\bf iii) General case of $C\neq 0$} 

Here we consider the metric of general case of $C\neq 0$, namely (\ref{10dmetric-2}) with (\ref{sol-10})-(\ref{sol-12}).
{In this case, we find $\bar{A}\neq \bar{n}$ since $\bar{A}$
and $\bar{n}$ are modified by the term $\tilde{c}$ , and $\bar{n}$ could have a zero point
as in the case of the black hole configuration when $\tilde{c}$ exceeds a critical value. Then
we could find a phase transition by adding the term $C$ to the solution with AdS$_4$
boundary \cite{EGR}. 

In the confinement phase, we find the position of the domain-wall is pushed to $r=0$
by increasing $\tilde{c}$. In the case of dS$_4$ boundary, the horison is pushe toward
to larger $r$.
In any case, the boundary metric at $r=0$ is deformed. This point is seen 
as follows.
}
The five dimensional 
part of the metric is rewritten in terms of ${z^*}=1/z^2$ as follows
\beq
    ds^2_{(5)}=R^2\left({1\over z^*}\hat{g}_{\mu\nu}dx^{\mu}dx^{\nu}+{d{z^*}^2\over 4{z^*}^2}\right)\, ,
      \label{IR-metric} 
\eeq
and the 4D part is expanded by the powers of $z$ for $R=1$, as follows
\beq
  \hat{g}_{\mu\nu}=\hat{g}_{(0)\mu\nu}+\hat{g}_{(2)\mu\nu}z^*+
 {{z^*}^2}\left(\hat{g}_{(4)\mu\nu}+\hat{h}_{1(4)\mu\nu}\log z^*
+\hat{h}_{2(4)\mu\nu}(\log z^*)^2\right)+\cdots\, .
\eeq\label{Feff1IR}
The first term is given as 
\bea
 \hat{g}_{(0)\mu\nu}&=&(\hat{g}_{(0)00},~\hat{g}_{(0)ij})\, \nonumber \\
         &=&\left(-{\left(({b}{b}_1)^2-\tilde{c}_0\right)^2 \over \left({b}^4+\tilde{c}_0\right)r_0^4},
       ~{{b}^4+\tilde{c}_0\over r_0^4}a_0(t)^2\gamma_{i,j}\right)\, , \label{Feff2IR}    
\eea
where
\beq\label{Feff3IR} 
   {b}^2=-{\lambda\over 4}R^4\, , \quad {b}_1^2=-{\lambda+\dot{\lambda}a_0/\dot{a}_0\over 4}R^4\,
\eeq
This implies that the boundary metric depends on the dark 
{radiation, namely} the SYM fields. Then this fact seems to be
contradict with the expectation of the
decoupling of SYM theory and the gravity on the boundary. 
On the other hand, 
at the $r=\infty$, the boundary metric is not affected by this dark radiation $C$ as expected.
Then the holographic situation is modified in the side of $r=0$ for $C\neq 0$, where the dual 4D theory
couples with gravity through the energy momentum tensor{ \footnote{We could show that the 
vacuum expectation value of the energy momentum tenser in the IR side boundary is also derived according to
the renormalization group method used in the UV side. The result at IR side is given by the same form of
the one at UV side by replacing the curvatures written by the metric (\ref{Feff1IR}).
For example, the trace anomaly is given by $\langle T_{\mu}^{\mu}\rangle={N^2\over 32\pi^2}\left(
       R^{\mu\nu}R_{\mu\nu}-{1\over 3}R^2\right)$. See Appendix B.}}
generated in the side of SYM theory. It is an interesting
problem to make clear this point and to investigate the holography of a {SYM} theory coupled with the gravity.
We however postpone to investigate this problem
to the future, and we restrict to the case of $C=0$ hereafter.

\vspace{.3cm}
\section{Gravity dual and domain wall}

We consider the gravity dual of the {two theories} living 
on different boundaries. It is represented by 
a common bulk manifold. We study how we can see the holographic properties 
of two field theories on the same bulk manifold through various objects 
which are responsible {for} field theories. 

\vspace{.3cm}
\subsection{Wilson-Loop and Quark Confinement}

The potential between
quark and anti-quark is studied by the Wilson-Loop. It is obtained holographically
from the U-shaped ( in $r-x$ plane) string which is embedded in the bulk
and its two end-points are on 
the boundary. 
Supposing a string whose world volume is set in $(t,x)$ plane 
\footnote{Here $x$ denotes one of the three coordinate $x^i$, and we take $x^1$ in the present case.}, 
the energy $E$ 
of this state is obtained as a function of the distance ($L$) between
the qaurk and anti-quark according to \cite{GIN1}.

Taking the gauge as $X^0=t=\tau$ and $X^1=x^1=\sigma$
for the coordinates $(\tau,~\sigma)$ of string world-volume,
the Nambu-Goto Lagrangian in the present background (\ref{10dmetric-2}) 
becomes
\beq
   L_{\textrm{\scriptsize NG}}=-{1 \over 2 \pi \alpha'}\int d\sigma ~
   {\bar{n}(r)}\sqrt{r'{}^2
        +\left({r\over R}\right)^4 \left({\bar{A}(r)}a_0(t)\gamma(x)\right)^2 } ,
 \label{ng}
\eeq
where
\beq
  \gamma(x)={1\over 1-x^2/4}\, ,
\eeq
and we notice $r'=\partial r/\partial x=\partial r/\partial \sigma$. and 
{$\bar{n}, \bar{A}$
have no time dependence since we here use the metric (\ref{AdS4-2}).}
The general form of the solution for $\bar{n}$ and $\bar{A}$ are given in (\ref{sol-10})-(\ref{sol-12}), and
they depend on the time through the scale factor $a_0(t)$. Then, in order to see the static
energy as $E=-L_{\textrm{\scriptsize NG}}$, we should
restrict the solutions to the case of $C=0$ and $\lambda=-\lambda_0$.
 In this case, the two boundaries have the same form of metric as given in (\ref{AdS4-1})-(\ref{AdS4-3}) and
(\ref{AdS4-4-In}) with different notation of the radial coordinate for each boundary, $r$ and $z$ respectively.

\vspace{.5cm}
In the case of (\ref{AdS4-1})-(\ref{AdS4-3}), the energy is rewritten to a more convenient form 
{by introducing the factor $n_s$ (given below)} \cite{Gub} as
\beq
 E=-L_{\rm NG}
  ={1\over 2\pi \alpha'} \int d \tilde{\sigma} ~ n_s ~
        \sqrt{1+ \left({R^2\over r^2 \bar{A}}\partial_{\tilde{\sigma}}r
               \right)^2}\ , \label{W-energy}
\eeq
\beq
  \tilde{\sigma}=a_0(t)\int d\sigma\gamma(\sigma)
       =a_0(t)\int d\sigma {1\over 1-\sigma^2/4}\, ,
\eeq
\beq 
 n_s=\left({r \over R}\right)^2 \bar{A}\bar{n}=\left({r \over R}\right)^2 \left(1+{r_0 ^2\over r^2}\right)^2 \, , \label{matter-ns0} 
\eeq
Here, we use
the proper coordinate $\tilde{\sigma}$ instead of the comoving coordinate ${\sigma}$ 
to measure the distance between the quark and anti-quark. 

\vspace{.5cm}
In this form, the criterion of the confinement is stated such that $n_s$ has a finite minimum value at some appropriate
$r(=r^*)$. In the present case, we find $r^*=r_0$. Actually, in such a case, $E$ is approximated as \cite{GIN1} 
\beq
 E\sim {n_s(r^{*})\over 2\pi \alpha'} {L}\ , \label{linear-P}
\eeq 
where 
\beq
  {L}=2\int_{\tilde{\sigma}_{min}}^{\tilde{\sigma}_{max}}d \tilde{\sigma}\, ,
\eeq
and $\tilde{\sigma}_{min}$ ($\tilde{\sigma}_{max}$) is the value at
$r_{min}$ ($r=\infty $) of the string configuration \cite{GIN2}.
The tension of the linear potential between the quark
and anti-quark is therefore given as
\beq \label{tension2}
 \tau_{q\bar{q}}={n_s(r_0)\over 2\pi \alpha'}\, .
\eeq
We notice that the U-shaped string configuration whose bottom point is near $r_0$ and the string on both sides
goes up toward the boundary $r=\infty$. When the bottom approaches to $r_0$ the length $L$ goes to $\infty$.
In other words, the string configuration is bounded at $r=r_0$ and cannot exceed this point to smaller $r$.

\vspace{.5cm}
In the case of (\ref{AdS4-4-In}), the procedure of the calculation of the Wilson loop is completely
parallel to the above case only by replacing $r$
by $z$. Then we find the same tension of the linear potential between the quark
and anti-quark, which are living on the boundary $r=0$ or $z=\infty$, is obtained as
\beq \label{tension3}
 \tau_{q\bar{q}}={n_s(r_0)\over 2\pi \alpha'}\, .
\eeq
{In the present case,  the string on both sides
goes up toward the boundary, namely to $r=0$.}  So
the U-shaped configuration of the string has a form which has been 
upside-down the one obtained
above. Then we will find two types of string configurations which 
are responsible to the
Wilson loop calculation. The end points of the one type of string 
go towards $r=\infty$, and the one of the other type goes to $r=0$. 
This equation is very complicated, so we show its numerical result
in the Fig.{\ref{wilson-loop}}

\vspace{.5cm}
\noindent{\bf String configurations and Domain Wall}

\begin{figure}[htbp]
\vspace{.3cm}
\begin{center}
\includegraphics[width=14.0cm,height=7cm]{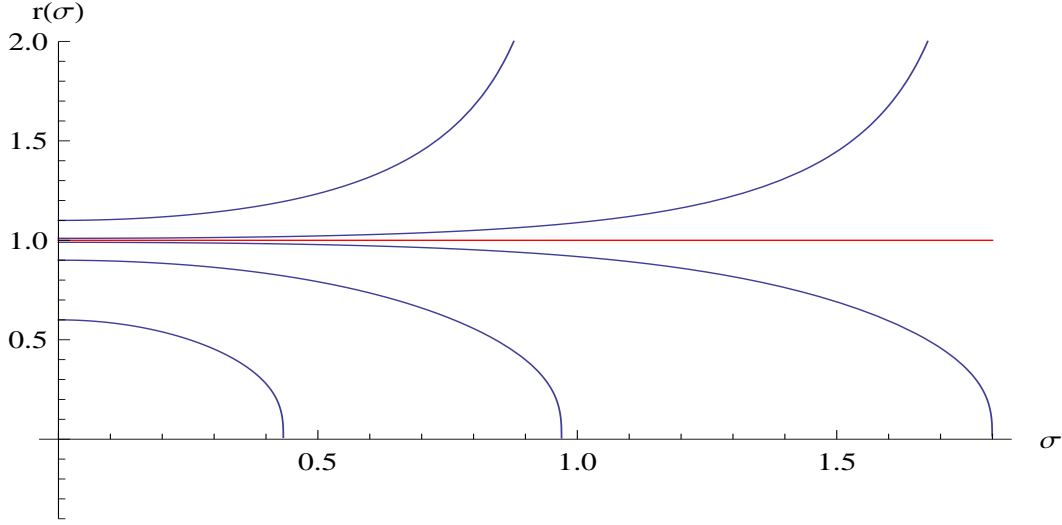}
\caption{Solutions for $r(\sigma)$, where $\tilde{\sigma}$ is denoted by $\sigma$. The curves denote
for $r(0)=1.1,~1.01,~0.99,~0.9,~0.6$ from the above one. The horizontal line $r=1.0$ shows the domain wall.
\label{wilson-loop}}
\end{center}
\end{figure}

Here we show the string configurations mentioned above to make clear the situation. They are obtained by solving the 
equation of motion for the profile of the string. Both the solutions belonging to the boundary $r=\infty$ and
$r=0$ are obtained by solving the same equation, which is given from (\ref{ng}) as follows
\beq\label{string-eq}
  r''-r'\left( \log\left( {r^4\over R^4}+\left({r'\over (1+(r_0/r)^2)}\right)^2\right)\right)'
  +2{r_0^3\over r^3}{{r'}^2\over 1+(r_0/r)^2}-2{r^3\over R^4}\left(1-\left({r_0\over r}\right)^4\right)=0\, 
\eeq
where prime denote the differation with respect to $\tilde{\sigma}$ as $r'=\partial_{\tilde{\sigma}}r$.
This equation is complicated, so we solve it numerically,

Several configurations are shown in the Fig.\ref{wilson-loop}, from which we can see the solutions are
separated to two groups by the boundary condition at $\sigma=x=0$, namely the value of $r(0)$. The wall which
separate two classes of the solutions is found at $r(0)=r_0$.

\vspace{.5cm}
\subsection{D7 brane embedding and domain wall}

Here, we study the D7 brane embedding, which is responsible for studying the meson spectrum
and the chiral condensate of the boundary theory.
The D7-brane action is given by the Dirac-Born-Infeld (DBI) and the
Chern-Simons (CS) terms as follows, 
\bea\label{d7action}
S_{D7}&=&-T_{7}\int d^8\xi
 e^{-\Phi}\sqrt{-\det\left(g_{ab}+2\pi\ap F_{ab}\right)}+T_{7}\int \sum_i
\left(e^{2\pi\ap F_{(2)}}\wedge c_{(a_1\ldots
a_i)}\right)_{0\ldots 7}~,\\
g_{ab}&\equiv&\p_a X^{\mu}\p_b X^{\nu}G_{\mu\nu}~, \qquad
c_{a_1\ldots
a_i}\,\equiv\,\p_{a_1}X^{\mu_1}\ldots\p_{a_i}X^{\mu_i}C_{\mu_1\ldots\mu_i}~.\nonumber
\eea
where $T_7$ is the brane tension.  
The DBI action involves
the induced metric $g_{ab}$ and the $U(1)$ world volume
field strength $F_{(2)}=d A_{(1)}$.

\vspace{.6cm}
\noindent{\bf Near $r=\infty$}
\vspace{.3cm}

For the simplicity, we consider the background
(\ref{AdS4-1}) - (\ref{AdS4-2}) near the boundary $r=\infty$.  The metric 
{of the} extra six dimensional part of this metric 
is rewritten as follows
\beq
   \frac{R^2}{r^2} dr^2 +R^2d\Omega_5^2=\frac{R^2}{r^2}\left(
       d\rho^2+\rho^2 d\Omega_3^2+\sum_{i=8}^9{dX^i}^2\right)\, ,
\eeq
where the new coordinate $\rho$ is introduced instead of $r$ with the relation
\beq
  r^2=\rho^2+(X^8)^2+(X^9)^2
\eeq
Thus, the induced metric of the D7 brane is obtained as
\beq\label{8dmetric}
ds^2_{8}={r^2 \over R^2}\left(1+{r_0^2\over r^2}\right)^2\left(-dt^2+a_0^2(t)\gamma^2(x)(dx^i)^2\right)+
\frac{R^2}{r^2} \left(
       (1+{w'}^2)d\rho^2+\rho^2 d\Omega_3^2 \right) \, , 
\label{finite-c-sol-3}
\eeq
where the profile of the D7 brane is taken as $(X^8,X^9)=(w(\rho),0)$
and $w'=\partial_{\rho}w$, then 
\beq
  r^2=\rho^2+{w}^2\, .
\eeq
In the present case,
there is no R-R filed, so the action is given only by the one of DBI as
\beq
  S_{D7}=-T_{7}\Omega_3\int d^4x a_0^3(t)\gamma^3(x)\int d\rho \rho^3
   \bar{A}^4\sqrt{1+{w'}^2(\rho)}\, ,
\eeq 
where
\beq
  \bar{A}=\left(1+{r_0^2\over r^2}\right)^2
\eeq 
and $\Omega_3$ denotes the volume of $S^3$ of the D7's world volume.
\begin{figure}[htbp]
\vspace{.3cm}
\begin{center}
\includegraphics[width=14.0cm,height=7cm]{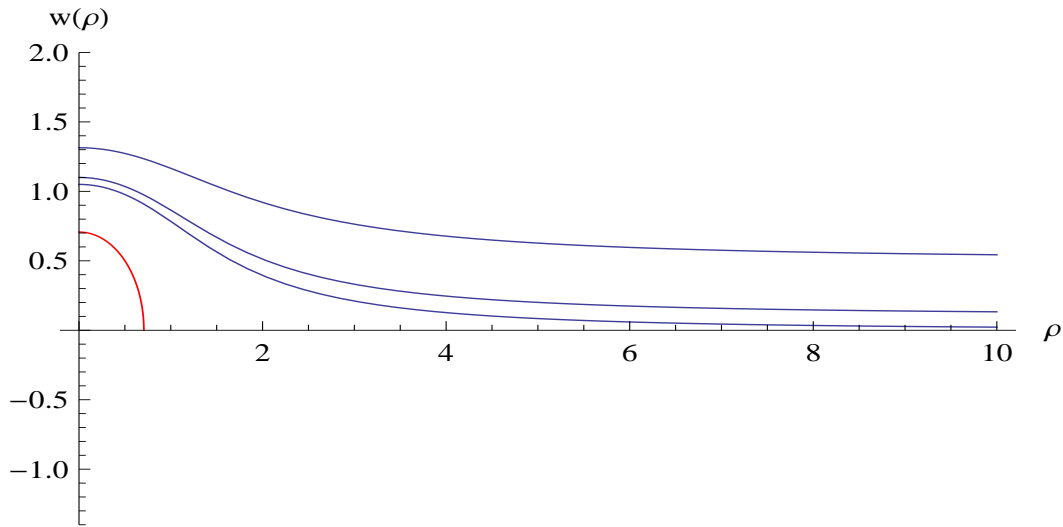}
\caption{Typical solutions of $w(\rho)$ for $\lambda_0=2, ~\mu=1/R=1.0$, $r_0=1/\sqrt{2}$. 
The curves are given for $w(0)=1.3,~1.1,~1.05$ from the above one to the below. 
The circle represents $r=\sqrt{r_0^2-\rho^2}$, which corresponds the domain wall of the
dual bulk manifold for two boundaries.
\label{chiral-c}}
\end{center}
\end{figure}

\vspace{.5cm}
From this action, the equation of motion for $w$ is obtained as
\beq\label{eq-w}
  w''+\left({3\over \rho}+{\rho+ww'\over r}\partial_r (\log (\bar{A}^4))\right)
     w'(1+{w'}^2)-{w\over r}(1+{w'}^2)^2\partial_r ( \log (\bar{A}^4) )=0\, .
\eeq
The constant $w$ is not the solution of this equation, so the supersymmetry is broken.
The numerical solutions of (\ref{eq-w}) for $w(\rho)$ are shown in the Fig. \ref{chiral-c}.
{In general, in this case, we find finite chiral condensate $\langle\bar{\Psi}\Psi\rangle=c$ 
for any $m_q\geq 0$ since 
the curves decrease from the above with increasing $\rho$.
For all curves, we find the behavior given by the following asymptotic form}
\beq\label{chiral-br}
  w=m_q+{c+4m_q^2r_0^2\log (\rho)\over \rho^2}+\cdots\, ,
\eeq
at large $\rho$ with $c>0$. Here, the term proportional to $\log (\rho)$ comes from the 
breaking of the conformal invariance due to the cosmological constant in the theory \cite{H,GIN1,GIN2}.
We can {observe} spontaneous chiral symmetry breaking from the third curve, which
corresponds to $m_q=0$. It shows the mass generation of a massless quark due to the
chiral condensate $\langle\bar{\Psi}\Psi\rangle$. 

As a result, we could say that the spontaneous mass generation of massless quarks
is realized in the theory on the boundary at $r=\infty$.
{This} point is already found previously \cite{GN13}. 
We notice here that the embedded region of D7 brane with $m_q\geq 0$ is restricted to
the region $r> r_0$. {Furthermore, there is no D7 brane configuration which crosses the domain wall
$r=r_0$ in $w-\rho$ plane.}
Then the quarks introduced in the dual SYM theory on the boundary $r=\infty$
can be represented by the D7 brane embedded in the region of $r>r_0$.

\vspace{.6cm}
\noindent{\bf Near $r=0$}
\vspace{.3cm}

For the flavor brane near $r=0$, its embedding is performed as follows. First, 
by adopting the bulk metric (\ref{AdS4-4-In}), the procedure is completely parallel to
the above case by replacing $r$ by $z$. Then the embedded D7 branes of $m_q\geq 0$ are all
obtained in the region of $z>r_0$ and we find the dual theory with chiral symmetry breaking phase
at  the boundary $r=0$. In the present case, the region $z>r_0$ means $r<r_0$ since $z=r_0^2/r$.
{Then we find the fact that each theory in two boundaries 
of the bulk is separated by the wall at $r=r_0$.  }
Namely, we can study each dual theories can be given by considering the gravity within each region.

\vspace{.5cm}
\subsection{D5 Branes and Baryon}

Next, we consider the baryon. It is constructed from a vertex and $N_c$ quarks, and the latter are expressed 
by fundamental strings. The vertex is identified with the D5 brane, which is embedded
in the bulk as a probe with a non-trivial $U(1)$ flux in it. Then a baryon is discussed through the D5 brane
embedding given as follows.

First, we briefly review the model based on type IIB superstring theory \cite{wittenbaryon,imamura,cgs,cgst,GI}.
In the type IIB model, the vertex is described
by the D5 brane which wraps $S^5$ of the 10D manifold $M_5\times S^5$. In this case, in the bulk, there exists 
the following form of self-dual Ramond-Ramond field strength
\bea\label{fiveform}
G_{(5)}&\equiv &dC_{(4)}={4\over R}\left(\epsilon_{S^5}+{}^*\epsilon_{S^5}\right)\, \\
            \epsilon_{S^5} &=&R^5\,\mbox{vol}(S^5) d\theta_1\wedge\ldots\wedge d\theta_5
\eea 
where $\mbox{vol}(S^5)\equiv\sin^4\theta_1
\mbox{vol}(S^4)\equiv\sin^4\theta_1\sin^3\theta_2\sin^2\theta_3\sin\theta_4$,
and  $\epsilon_{S^5}$ denotes the volume form of $S^5$ part. 
The flux from the stacked D3 branes flows into the D5 brane
as $U(1)$ field which is living in the D5 brane. 

\vspace{.5cm}
The effective action of D5 brane is given by using the Born-Infeld and
Chern-Simons term as follows
\begin{eqnarray}\label{d5action}
S_{D5}&=&-T_{5}\int d^6\xi
 e^{-\Phi}\sqrt{-\det\left(g_{ab}+2\pi\ap F_{ab}\right)}+T_{5}\int
\left(2\pi\ap F_{(2)}\wedge c_{(4)}\right)_{0\ldots 5}~,\\
g_{ab}&\equiv&\p_a X^{\mu}\p_b X^{\nu}G_{\mu\nu}~, \qquad
c_{a_1\ldots
a_4}\,\equiv\,\p_{a_1}X^{\mu_1}\ldots\p_{a_4}X^{\mu_4}C_{\mu_1\ldots\mu_4}~.\nonumber
\end{eqnarray}
where $T_5=1/(\gs(2\pi)^{5}\ls^{6})$ and  $F_{(2)}=d A_{(1)}$, which represents the $U(1)$ worldvolume
field strength.
In terms of (the pullback of)
the background five-form field strength $G_{(5)}$, the above action can be rewritten as
$$
S_{D5} = -T_5 \int d^6\xi~ e^{-\Phi}
     \sqrt{-\det(g+F)} +T_5 \int A_{(1)}\wedge G_{(5)}~,
$$

The embedding of the D5 brane is performed by solving the $r(\theta)$, $x(\theta)$, and $A_{(1)}(\theta)$
\cite{GI}.
They are retained as dynamical fields in the D5 brane action as the function of $\theta\equiv\theta_1$ only. 
The equation of motion for the gauge field $A_{(1)}$ is written as
$$
\partial_\theta D = -4 \sin^4\theta,
$$
where the dimensionless displacement 
is defined as the variation of the
action with respect to $E=F_{t\theta}$, namely
$D=\delta \tilde{S}/\delta F_{t\theta}$ and 
$\tilde{S}=S/T_5 \Omega_{4}R^4$. The solution to this equation 
is
\beq \label{d}
D\equiv D(\nu,\theta) = \left[{3\over 2}(\nu\pi-\theta)
  +{3\over 2}\sin\theta\cos\theta+\sin^{3}\theta\cos\theta\right].
\eeq
Here, the integration constant $\nu$ is expressed as $0\leq\nu=k/N_c\leq 1$,
where $k$ denotes the number of Born-Infeld strings emerging from one of the pole of
the ${\bf S}^{5}$. 

\vspace{.5cm}
Next, it is convenient to eliminate the gauge 
field
in favor of $D$, then the Legendre transformation is performed for the original Lagrangian to
obtain an energy
functional as \cite{cgs,cgst,GI}:
\beq \label{u}
U = {N\over 3\pi^2\alpha'}\int d\theta~\bar{n}
\sqrt{r^2+r^{\prime 2} +(r/R)^{4}x^{\prime 2}(\bar{A}a_0\gamma)^2}\,
\sqrt{V_{\nu}(\theta)}~.
\eeq
\beq\label{PotentialV}
V_{\nu}(\theta)=D(\nu,\theta)^2+\sin^8\theta
\eeq
where we used $T_5 \Omega_{4}R^4=N/(3\pi^2\alpha')$, and we use the metric form (\ref{10dmetric-2}).
Then, in this expression, (\ref{u}), 
$r(\theta)$ and $x(\theta)$ are remained, and they are solved by minimizing $U$. As a result,
the D5 brane configuration is determined. 

For the simplicity, here, we restrict to the point like configuration, namely $r$ and $x$ are constants.
Further, the simple metric (\ref{AdS4-2}) is adopted.
In this case, we have for the matter considered here
\beq \label{u-p}
U = ~r\bar{n}(r) U_0=r\left(1+{r_0^2 \over r^2}\right)U_0\, , 
\eeq
where $U_0$ is a constant given as
\beq \label{u-0}
  U_0= {N\over 3\pi^2\alpha'}\int d\theta \sqrt{V_{\nu}(\theta)}~.
\eeq
From (\ref{u-p}), we find that $U$ has a minimum at $r_m=r_0$. Then the vertex
is trapped at the domain wall.

We notice, however, the embedded region of the fundamental strings, quarks, are 
separated to two regions by the domain wall. Namely, the strings cannot cross 
the domain wall. In this sense, the baryons {are also} separated to two theories by the wall
in the gravity side.

\vspace{.3cm}
\section{Entanglement Entropy and domain wall}

Next, we consider the entanglement entropy for the theory of one boundary.
It is given by calculating the minimum area of 
the surface $A$ whose boundary $\partial A$ is set at the boundary
of the bulk and the surface could be extended in the bulk. In this calculation, there is a 
possibility that the minimal surface could penetrate into the bulk region corresponding to the 
theory living in the other boundary. When this situation
is realized, we could see a new entanglement
of two theories which are living in the separated boundaries.

{In order to see such phenomenon, we estimate
the entanglement entropy of a theory in one boundary according  
to the formula (3.3) in \cite{RT}
\beq\label{EEdef}
S_{EE}={{\rm Area(\gamma_A)}\over{4G_N^{(5)}}}\, ,
\eeq
where $\gamma_A$ denotes the minimal  surface,
whose boundary is defined by $\partial A$ and the surface is extended into the bulk.
And $G_N^{(5)}=G_N^{(10)}/(\pi^3R^5)$ 
denotes the 5D newton constant reduced from the 10D one  $G_N^{(10)}$. If, 
in this calculation, the minimal surface
crosses the domain wall, then we can say that a new kind of
an entanglement between two theories on each boundary may
exist. This is because the surface or equivalently 
the entanglement entropy is controlled by the dynamics of the other theory.

We adopt (\ref{AdS4-4-In}) as the bulk metric, which is given as
\beq\label{AdS4-4-2}
ds^2_{10}={z^2 \over R^2}\left(1+{r_0^2\over z^2}\right)^2 ds^2_{FRW_4}+
\frac{R^2}{z^2} dz^2 +R^2d\Omega_5^2 \ . 
\eeq
where
\beq
  ds^2_{FRW_4}=-dt^2+a_0^2(t)\gamma^2 \left(dp^2+p^2 d\Omega_2^2\right)\, , \label{AdS4-4-3}
\eeq
\beq
 p={\bar{r}\over \bar{r}_0}\, , \quad  \gamma=1/ (1-p^2/4) \, ,
\eeq
We notice here that the mass dimension is -1 for $a_0$, but $p$ has no mass dimension since it is scaled
by $\bar{r}_0$.
In order to study the entanglement entropy (EE), we separate the 3D {space} of the boundary at a fixed time
by a constant value for $p=p_0$. Then the EE for the restricted space $p<p_0$ is obtained holographically
by finding the minimum value of the following quantity
\bea
  {S_{\rm area}\over 4\pi}&=&\left(  {r_0\over R}a_0\right)^3
  {\int^{\epsilon}_{x_0}} dx f_1(x) \, \label{EE-1} \\
       f_1(x)&=& p^2\left( {1+x^2\over x}\right)^3\gamma^3\sqrt{ {p'}^2+{b\over \gamma^2 (1+x^2)^2}}\, \label{EE-2} \\
       b&=& {R^4\over a_0^2 r_0^4}\, , \quad p'={\partial p \over \partial x}\, , \label{EE-11}
\eea
where $x=z/r_0$, and $x_0$ denotes the end point of the embedded surface. Since this integral diverges at the ultraviolet
(UV) limit, the UV cutoff $\epsilon$ is introduced.
This represents the minimal surface of the ball embedded in the bulk.
\begin{figure}[htbp]
\vspace{.3cm}
\begin{center}
\includegraphics[width=14.0cm,height=7cm]{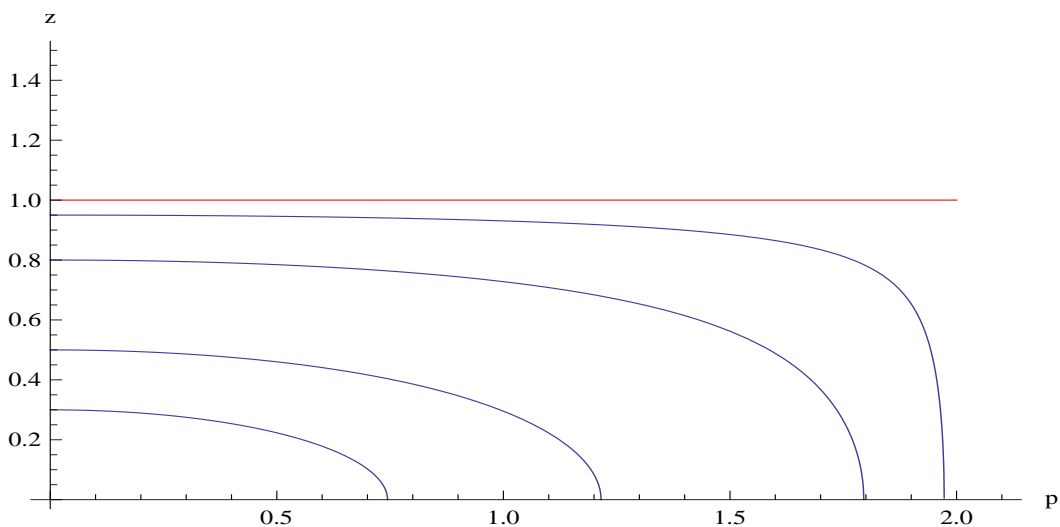}
\caption{Embedded solutions of $p(z)$ for $p_0=$ 0.74, 1.21, 1.79,  and 1.97 from the below. Other parameters
are set as {$r_0$=1, $R$=1} and $a_0$=0.4. 
The horizontal line represents the domain wall.
\label{z-p}}
\end{center}
\end{figure}

\vspace{.5cm}
In order to obtain the {minimum} of $S_{\rm area}$, we must 
solve the {variational equation} for $p(x)$ which
is extended in the region $0<x<x_0$ of the bulk space. On the other hand, the {information} of the two boundary
theories {is divided} by the domain wall as mentioned {above}. Then we will see the upperbound of $x_0$ at
$x_0=1$. This point corresponds to $r=r_0$. This is actually assured by rewriting the above $S_{\rm area}$ as
follows,\footnote{Notice that $p$ in Eq.(\ref{EE-4}) is a function of $y$ as solved from Eq.(\ref{EE-42}).}
\bea
   {S_{\rm area}\over 4\pi}&=&\left(  {r_0\over R}a_0\right)^3\int dy f_2(y) \, \label{EE-3} \\
       f_2(y)&=& \left( {1+x^2\over x}\right)^3\sqrt{1 +{b{(x'p^2\gamma^2)}^2\over (1+x^2)^2}}\, \label{EE-4} \\
       y&=&y(p)=\int^p dp {p^2\over (1-p^2/4)^3}\, , \quad x'=\partial_y x\, . \label{EE-42} 
\eea
These formula are very similar to the case of the Wilson loop calculation, where the embedded string configuration
is obtained as a U-shaped one, and its bottom point is bounded at the minimum of the prefactor of the
integrand. It {corresponds} here to
\beq
  n_{Sphere}=\left( {1+x^2\over x}\right)^3\, . \label{EE-5}
\eeq
In fact we can see that $n_{Sphere}$ has a minimum at $x=1$.  Then the embedded solution of the ball
would be bounded in the region $0<x<1$, and this is also assured from the numerical calculation as
shown in the Fig.\ref{z-p}.

The analysis given above is obtained for $p(z)$ with various $p_0$, which is the value of $p$
at the UV limit $z=0$. The bottom point of $p(z)$ approaches to the value for $z=r_0$ (the horizontal line $x=1$). {However,}
it does not never exceed this line. In other words, the quantum information of the theory on the other boundary
does not affect on the EE calculation of the theory at $z=0$. This implies that there is no entanglement among the 
two theories on the opposite boundaries at $r=0 (z=\infty)$ and $r=\infty (z=0)$.

\vspace{1cm}
\noindent{\bf Divergent term and central charge of the theory}

\vspace{.2cm}
While it is difficult to find an analytic solution of $p(z)$ in the present case, it is possible 
{to see the divergent form of ${S_{\rm area}\over 4\pi}$ near the UV limit by using the {approximate} solution 
near the boundary. Before solving
our present case, this point is shown firstly for the bulk AdS$_5$ case with Minkowski boundary metric.}
Through this analysis, we could obtain a knowledge {related to} the central charge of the theory. 
We write the AdS$_5$ metric as,
\beq
  ds^2_{\rm AdS5}=\left({R\over z}\right)^2\left(\eta_{\mu\nu}dx^{\mu}dx^{\nu}+dz^2\right)\, .
\eeq
In this case also, we use the {same} notation $p$ 
for the theree space radial coordinate as 
\beq
  ds^2_{(4)}=\eta_{\mu\nu}dx^{\mu}dx^{\nu}=-dt^2+dp^2+p^2d\Omega_{(2)}^2\, .
\eeq
Then {the 3D area of} the embedded ball with radius $p_0$ 
is given as{
\beq
   {S_{\rm area}\over{4\pi}}=R^3 \int_{z_{\rm min}}^{z_{\rm max}} {dz\over z^3} 
                   p^2\sqrt{1 +{p'}^2}\, \label{AdSEE-1} \, ,
\eeq
where $p'=\partial p/\partial z$ and $\epsilon$ is the cutoff, 
namely $z>\epsilon$.  In order to get the area of minimal surface, 
we should minimize $S_{\rm area}$.  This requirement is achieved by 
the variational principle.  
The variational equation for $p(z)$ is solved in this case as
\beq
  p=p_0 + p_2 z^2+p_4 z^4 + \cdots \, . \label{AdSEE-2}
\eeq
In the above, $p_0$ and $p_4$ are two arbitrary constants.  
This is the general solution.  Simple power counting assures that 
only $p_0$ and $p_2$ are necessary to get divergent terms of $S_{\rm area}$.  
The value of $p_2$ is determined as  
\beq
p_2=-{1\over{2p_0}}\, .  \label{p2}
\eeq
}
Using this solution, we can estimate the leading UV ($\epsilon\to 0$) 
divergent terms as
\beq\label{AdSEE-3}
  {{S_{\rm area}}\over{4\pi}}={1\over 2}R^3\left[\left( {{p_0^2}\over{\epsilon^2}}\right)+\log\left({ {{\epsilon}\over{p_0}} }\right)\right]+{\rm finite~ terms}\, ,
\eeq
where the parameter $p_0$, which characterize the present physical system, 
is introduced according to \cite{RT, RT2}.  

{The entanglement entropy is then expressed in the form used 
in \cite{RT2} as follows
\beq\label{EE2CC}
S_{EE}={{\gamma_1}\over 2}\cdot{{{\rm Area}(\partial A)}\over{4\pi \epsilon^2}}
+\gamma_2\log\left(p_0\over{\epsilon}\right)+{\rm finite~ terms}\, ,
\eeq
where  ${{\rm Area}(\partial A)}$ denotes the area of the surface $A$, and $\gamma_1$ and $\gamma_2$ 
are numerical constants.  
In the present case, ${{\rm Area}(\partial A)}=4\pi p_0^2$ , then
the coefficients $\gamma_1$ and $\gamma_2$ are obtained as 
\beq\label{ccAdS5}
{\gamma_1\over 2}={{2\pi R^3}\over{4 G_N^{(5)}}}=N^2\, ,\qquad
\gamma_2={{2\pi R^3}\over{4 G_N^{(5)}}}=N^2\, ,
\eeq
with the use of (\ref{EEdef}) and 
the relation $R^4=4\pi g_s{\alpha'}^2 N$.  
The result 
is  compared to the corresponding divergent terms of our AdS$_4$ space model in the following.}  
 
\vspace{.5cm}
Now we return to (\ref{EE-1}).  
In this case,  the variational equation is solved by using the following expansion, 
\beq
  p=p_0+p_2x^2+p_{4L}x^4\log x +p_4 x^4+\cdots \,  . 
\eeq
The coefficients of this series expansion are determined by the
two arbitrary constants, $p_0$ and $p_4$.  
{The value of $p_2$ and $p_{4L}$ are determined as 
\beq
p_2=-{{(1 - (p_0^2/4)^2) R^4}\over{2 a_0^2 p_0 r_0^2}}\, .\label{radsol-1}
\eeq
and 
\beq
p_{4L}=-\frac{\left(1 - \left(\frac{p_0^2}{4}\right)^2\right) R^8\cos ^2(\sqrt{\lambda_0}t)}{4 a_0^4 p_0 r_0^8} .\label{p4l}
\eeq}

This solution is not 
analytical in contrast to (\ref{AdSEE-2}) due to the term $\log x$.  However it is not 
important since only $p_0$ and $p_2$ contribute to the divergent terms of $S_{EE}$  
as is mentioned above.  
The value of $p_2$ is determined as 
\beq
p_2=-{{(1 - (p_0^2/4)^2) R^4}\over{2 a_0^2 p_0 r_0^2}}\, .\label{radsol-1}
\eeq
With this solution, we obtain 
\beq\label{EERW-1}
  {{S_{\rm area}}\over{4\pi}}={1\over 2}R^3\left[\left( {{k_0^2}\over{4\epsilon^2}}\sin^2(\sqrt{\lambda_0}t) \right)
 +(1+k_0^2\cos^2 (\sqrt{\lambda_0}t ) )\log\left({{\epsilon}\over{p_0}}\right)\right]+{\rm finite~ terms}\, ,
\eeq
where $k_0=p_0/(1-p_0^2/4)$.  This result is  
also written in the form of (\ref{EE2CC}) with the following
coefficients $\gamma_i$, 
\bea
{\gamma_1\over 2}&=&{\lambda_0 \over 4}N^2\, \\  \label{ccfrw4-1}
\gamma_2&=&N^2\left(1+k_0^2\cos^2 (\sqrt{\lambda_0}t) \right)\, .  \label{ccfrw4-2}
\eea
where we used the following proper area in this case,
\beq
   {{\rm Area}(\partial A)}=4\pi k_0^2 a_0(t)^2\, .
\eeq

In this case $\gamma_1/2$ is slightly different by the factor ${\lambda_0 / 4}$ from the one of the 
AdS$_5$ case in (\ref{ccAdS5}).  However, this difference {can be} removed by redefinition of the cut off parameter
$\epsilon$. Then the remaining coefficient $N^2$ represents the freedom of the dual theory, and this is
consistent with the previous result that the central charge of the dual theory has been given by $N^2$
through the calculation of the energy momentum tenser holographically (see the Appendix A) \cite{GN13}.

On the other hand, we find a definite difference in $\gamma_2$.
This is understood from the fact that $\gamma_2$
depends on the curvatures in the 4D boundary and the extrinsic curvatures of the boundary $A$ in general \cite{RT2}.
{When {$t=(n+1/2)\pi/\sqrt{\lambda_0}$ ($n$ is an 
integer),} $p_{4L}=0$ as seen from
(\ref{p4l}).  In this case, $\gamma_2$ becomes the same with (\ref{ccAdS5}) and  
independent of $p_0$. Thus, there is a relation between $\gamma_2$ and 
$p_{4L}$. {More precisely, $p_{4L}$ and the extra term in 
$\gamma_2$ both contain the facor $\cos^2(\sqrt{\lambda_0}t)$.}  
While we are still considering about the physical interpretation of this 
relation, this is remained as an open question.  

It would be an interesting problem to assure that our result could coincide with the one given from the side of the dual
field theory, the SYM theory in the AdS$_4$ background, in order to see the validity
of the gauge/gravity corresponding for our present model. This is remained here as an open question.

\vspace{.5cm}
\section{Summary and Discussions}
In this paper, we have put forward an {extended} 
form of AdS$_5$/CFT$_4$ 
duality proposed in the previous paper \cite{GN13}, where AdS$_5$ is 
replaced by $\widetilde{AdS_5}$ whose boundary (in the ultraviolet side) 
is expressed by the FRW$_4$ space-time with finite 4D curvature.  
However the notation $\widetilde{AdS_5}$ might be 
missleading. {It is because one might consider that, 
in order to get the solution $\widetilde{AdS_5}$, the equation of motion
would be different from the one which leads to the ${AdS_5}$. Contrary to this
expectation, $\widetilde{AdS_5}$ is a solution of the same 5D 
Einstein equation which leads to the typical ${AdS_5}$. Then the two
solution are locally same with each other. On this point, we will discuss more in the
next chance.
}
  
Two cases of the boundary geometry, 
AdS$_4$ and dS$_4$, are possible for this FRW$_4$ depending 
on the sign of the
4D cosmological constant $\Lambda_4$. The parameter $\Lambda_4$
can be introduced as an arbitrary constant in the process of
solving the 5D Einstein equation with negative 5D cosmological
constant $\Lambda$, which comes from five form field strength and is 
independent of $\Lambda_4$.

Here we point out a new holographic feature of the $\widetilde{AdS_5}$ with AdS$_4$ boundary. In this case, we observe
second boundary in $\widetilde{AdS_5}$ at the opposite side of the fifth coordinate, namely at $r=0$ in addition to
the one at $r=\infty$.
This fact is in sharp contrast to the usual asymptotic AdS$_5$ case, in which the boundary appears only at $r=\infty$ and
the point $r=0$ is usually set as a horizon. 

This situation depends also on the other parameter $C$ in the general form of $\widetilde{AdS_5}$
given in (\ref{sol-10})-(\ref{sol-12}), where $C$ denotes the 
dark radiation. {While this term pushes the the domain wall to smaller $r$,
we could find the boundary at $r=0$ for small value of $\tilde{c}_0$.}
{However, }the geometry of the boundary at $r=0$ is generally different from a simple AdS$_4$ 
which is realized at $r=\infty$.
In the subsection 2.3, short discussion is given in the case of $C\neq 0$, 
in which it is pointed out that the metric of the IR boundary depends  
on the dark radiation or the SYM fields. On the other hand, at UV 
boundary the situation is different. Its geometry is not affected by 
the dark radiation. This point is interesting but we postpone to resolve this 
problem in the future.

Thus we restricted here to the case of $C=0$ and constant $\lambda$ in order to simplify
the problem of two boundaries discussed in this article.  
In this case, we find that the metric of the UV boundary takes 
the form (\ref{AdS4-2}) and the one of IR boundary can be read from 
(\ref{AdS4-4-In}).  They have the same form if $z$ was identified with 
$r$.  Of course, they are different, but they are related as $z=r_0^2/r$ 
and the point $r=r_0$ has an important holographic meaning in the bulk. In fact, we find
that this point corresponds to the domain wall.  

As assured from the metric (\ref{AdS4-4-In}), we could observe that the 4D
dual theory living at the boundary $r=0$ is also the SYM theory in the confinement phase. Furthermore,
from the scaling behavior of the metric form of IR boundary (\ref{AdS4-4-In}),
the limit of $r=0$ doesn't correspond to the IR but to the UV limit of the corresponding 4D theory.
Then there are two holographic screens in this case. This implies that the 
two field theories are described by a common gravity dual, $\widetilde{AdS_5}$ with AdS$_4$ boundary.
We notice that we find one boundary at $r=\infty$ and a horizon in the infrared side at finite $r$
for another case of $\widetilde{AdS_5}$, which has dS$_4$ boundary.

The problem in the case of two boundaries is
how the bulk manifold $\widetilde{AdS_5}$ would provide informations of the two field theories living on the 
different boundaries. Is it possible to get precise dynamical informations of two theories separately from
the common bulk geometry? We could show that the answer is yes for this question 
interms of the presence of a sharp domain wall
in the bulk. The gravity duals $\widetilde{AdS_5}$ for the two theories are separated by this wall.

The existence of the domain wall is assured by embedding the fundamental string, D7 brane and D5 brane in $\widetilde{AdS_5}$.
These objects give us the information of the Wilson-Loop, quarks, meson spectrum and baryons of the dual SYM theory. We could
find that the embedded regions of these objects are restricted to the either side of the bulk separated by the domain wall. In other words, these extended objects
cannot be embedded across the domain wall. Then the property of the field theory in one boundary is given by the gavity of one-side bulk devided by the domain wall.

Another interesting embedding problem is found in the calculation of the entanglement entropy $S_{\rm EE}$. This is obtained 
by the minimal surface ${\rm Area}(\gamma_A)$, which is defined as the minimum of the embedded surface whose boundary separates
the fixed-time boundary space into two regions. In this calculation, we find the embedded surface never extend across the domain
wall as other embedded stringy objects discussed above. 
This fact implies that the quantum fields in the theories of the two boundary don't affect each other. 
 
As for the entanglement entropy $S_{\rm EE}$ defined in either boundary. 
It diverges in general and written as (\ref{EE2CC}). The two coefficients
$\gamma_1$ and $\gamma_2$ of this expression reflect the freedom of the quantum fields of the theory and
the geometry of the 4D space-time of the boundary respectively. The result, ${\gamma_1\over 2}=N^2$, is common to the one of the case
of AdS$_5$ when we take the area of the sphere of the three space boundary by using the proper distance in the $\widetilde{AdS_5}$ case 
with AdS$_4$ boundary. On the other hand, $\gamma_2$ depends on the 4D curvatures and extrinsic curvatures on the 3D sphere. These quantities
largely changes $\gamma_2$ of $\widetilde{AdS_5}$ from the one of AdS$_5$. Our result, (\ref{ccAdS5}), for $\gamma_2$ would be important to assure 
the curvature dependence of $S_{\rm EE}$ in curved space-time. We will discuss this point in the future work. 

\vspace{.5cm}
{Finally we give the following two comments. First, the boundary of the AdS$_5$ is considered
here as the point where double pole (as given by Witten in \cite{MGW}) 
with respect to the fifth coordinate is observed. Two such points are found 
here at $r=0$ and $r=\infty$ for (A)dS$_4$ slice. 
Of course, another kinds of boundary can be considered
as discussed in  \cite{SN}. In \cite{SN}, the authors have examined the bulk fields
near a bulk singularity by supposing the existence of a new CFT there. 

As for the boundary as a double pole point, the double boundaries are also observed
in the black hole type solutions. In \cite{EM}, the so called
topological black hole solutions are discussed. While we don't consider this type geometry, 
we can see that
this case is similar to our solution of
dS$_4$ slice since a horizon exists between the two boundaries in both cases.

\vspace{.3cm}
Secondly, we should notice the following point. 
In \cite{EGR2}, it is shown that our solutions used here can be rewritten to
the form of the topological black hole solutions by a coordinate transformation. 
This is not surprising because
both solutions are obtained from the same bulk Einstein equations which are derived from 
the action of Einstein-Hilbert and 5D cosmological constant as mentioned above. 
However, this transformation
is performed in 5D by a kind of Rindler transformation,
then the slice of the 4D space-time and the fifth coordinate are 
changed. 
As a result, the properties of the CFT in the sliced 4D space-time are also changed. This point is
important and 
really assured by various holographic methods and quantities. So we think that the dual theory of the
topological black hole 
is different from our presnt case given in this article.
As mentioned in the first paragraph of this section, $\widetilde{AdS_5}$ is rewritten by
${AdS_5}$ through an appropriate coordinate transformation. However we should notice 
that we can see the properties of the CFT in 4D space-time, which is deformed from 4D
Minkowski space-time, through $\widetilde{AdS_5}$.
}

\newpage
\def\theequation{A.\arabic{equation}}
\setcounter{equation}{0}
\appendix

\noindent
{\Large {\bf Appendix}}
\section{ $\langle T_{\mu\nu}\rangle$ of the dual theory at $r=\infty$} \label{sec:dsl}

At first, we show 4D stress tensor of the boundary theory at $r=\infty$. Previously, it has
already given, so we review it briefly.  
First we rewrite
the 5d part of the metric (\ref{10dmetric-2})
according to the Fefferman-Graham framework \cite{KSS,BFS,FG}. Then it is given as
\bea
 ds^2_{(5)}&=&{1\over \rho}\left(-\bar{n}^2dt^2+\bar{A}^2a_0^2(t)\gamma^2(x)(dx^i)^2\right)+
{d\rho^2\over 4\rho^2}\, \label{thermal} \\
     &=&{1\over \rho}\hat{g}_{\mu\nu}dx^{\mu}dx^{\nu}+{d\rho^2\over 4\rho^2}\, ,
      \label{UV-metric} 
\eea
where $\rho=1/r^2$, $R=1$ and
\bea
 \bar{A}&=&\left(\left(1-{\lambda\over 4\mu^2}\left({\rho\over R^2}\right)\right)^2+\tilde{c}_0 \left({\rho\over R^2}\right)^{2}\right)^{1/2}\, , \\
\bar{n}&=&{\left(1-{\lambda\over 4\mu^2}\left({\rho\over R^2}\right)\right)
         \left(1-{\lambda+\dot{\lambda}a_0/\dot{a}_0\over 4\mu^2}\left({\rho\over R^2}\right)\right)-\tilde{c}_0 \left({\rho\over R^2}\right)^{2}\over  \bar{A}}
\eea
In the next, $\hat{g}_{\mu\nu}$ is expanded as \cite{BFS}
\beq
  \hat{g}_{\mu\nu}={g}_{(0)\mu\nu}+{g}_{(2)\mu\nu}\rho+
 {\rho^2}\left({g}_{(4)\mu\nu}+{h}_{1(4)\mu\nu}\log\rho
+{h}_{2(4)\mu\nu}(\log\rho)^2\right)+\cdots\, .
\eeq\label{Feff1}
where 
\beq
 {g}_{(0)\mu\nu}=({g}_{(0)00},~{g}_{(0)ij})=(-1,~a_0(t)^2\gamma_{i,j})\,  , \label{g00-1} 
\eeq
and 
\beq 
{g}_{(2)\mu\nu}={\lambda\over 2}\left(1+{{a_0\over \dot{a}_0}\dot{\lambda}\over \lambda},-~{g}_{(0)ij}) \right)\, ,  \label{g00-2} 
\eeq
\beq
 {g}_{(4)\mu\nu}={\tilde{c}_0\over R^4}~(3,~{g}_{(0)ij})+
                {\lambda^2\over 16}~\left(-{(\lambda+{a_0\over \dot{a}_0}\dot{\lambda})^2\over \lambda^2},~{g}_{(0)ij}\right)\, . \label{g00-3} 
\eeq\label{Feff2}

\vspace{.5cm}
Then by using the following formula \cite{KSS},
\beq\label{Feff-5}
 \langle T_{\mu\nu}\rangle={4R^3\over 16\pi G_N}\left({g}_{(4)\mu\nu}-
 {1\over 8}{g}_{(0)\mu\nu}\left( ({\rm Tr}g_{(2)})^2-{\rm Tr}g_{(2)}^2\right)
   -{1\over 2}\left({g}_{(2)}^2\right)_{\mu\nu}+{1\over 4}{g}_{(2)\mu\nu}
    {\rm Tr}g_{(2)}\right)\, ,
\eeq
we find 
\beq
 \langle T_{\mu\nu}\rangle=\langle \tilde{T}_{\mu\nu}^{(0)}\rangle+
{4R^3\over 16\pi G_N^{(5)}}\left\{{3\lambda^2\over 16}\left(1,~\beta{g}_{(0)ij}\right)\right\}\, .
\label{Feff6}
\eeq
\beq
   \langle \tilde{T}_{\mu\nu}^{(0)}\rangle={4R^3\over 16\pi G_N^{(5)}}
{\tilde{c}_0\over R^4}(3,~{g}_{(0)ij})\, , \quad
  \beta=-\left(1+{2{a_0\over \dot{a}_0}\dot{\lambda}\over 3\lambda}\right)\,\, . \label{Feff6-2}
\eeq
where $\langle \tilde{T}_{\mu\nu}^{(0)}\rangle$ comes from 
the conformal YM fields given in \cite{GN13}, so we find no anomaly for this component,
\beq
  \langle \tilde{T}_{\mu}^{{(0)}\mu}\rangle=0\, .\label{anomaly2-0}
\eeq
The second term corresponds to the loop corrections of the YM fields in the curved
space-time, and we find the conformal anomaly due to this term as
\beq
 \langle T_{\mu}^{\mu}\rangle=-{3\lambda^2\left(1+{\dot{\lambda}\over 2\lambda}{a_0\over \dot{a}_0}\right)\over 8\pi^2}N^2\, ,\label{anomaly2}
\eeq
where we used $G_N^{(5)}=8\pi^3{\alpha'}^4g_s/R^5$ and $R^4=4\pi N{\alpha'}^2g_s$.

\vspace{.3cm}
The above anomaly (\ref{anomaly2}) is obtained 
from the loop corrections of $\cal{N}=$ $4$ SYM theory
in a space-time, $g_{(0)\mu\nu}$, which is given by (\ref{g00-1}).
For this metric, the curvature squared terms responsible to the anomaly are given as
\bea
  R^{\mu\nu\lambda\sigma}R_{\mu\nu\lambda\sigma}&=&12\left(
      2\lambda^2+{\dot{\lambda}\lambda}{a_0\over \dot{a}_0}+\left(\dot{\lambda}{a_0\over 2\dot{a}_0}\right)^2\right)\, , \\
   R^{\mu\nu}R_{\mu\nu}&=&12\left(
      3\lambda^2+3{\dot{\lambda}\lambda}{a_0\over 2\dot{a}_0}+\left(\dot{\lambda}{a_0\over 2\dot{a}_0}\right)^2\right)\, , \\
   {1\over 3}R^{2}&=&12\left(
      4\lambda^2+4{\dot{\lambda}\lambda}{a_0\over 2\dot{a}_0}+\left(\dot{\lambda}{a_0\over 2\dot{a}_0}\right)^2\right)\, .
\eea
In general, the conformal anomaly for $n_s$ scalars,
$n_f$ Dirac fermions and $n_v$ vector fields is given as \cite{birrell,Duff93}
\beq
 \langle T_{\mu}^{\mu}\rangle=-{n_s+11n_f+62n_v\over 90\pi^2}E_{(4)}
        -{n_s+6n_f+12n_v\over 30\pi^2}I_{(4)}\, ,
\eeq
\beq
  E_{(4)}={1\over 64}\left(R^{\mu\nu\lambda\sigma}R_{\mu\nu\lambda\sigma}
       -4R^{\mu\nu}R_{\mu\nu}+R^2\right)\, ,
\eeq
\beq
  I_{(4)}=-{1\over 64}\left(R^{\mu\nu\lambda\sigma}R_{\mu\nu\lambda\sigma}
       -2R^{\mu\nu}R_{\mu\nu}+{1\over 3}R^2\right)\, ,
\eeq
where $\Box R$ has been abbreviated since it does not contribute here.
For the $\cal{N}=$ $4$ SYM theory, the numbers of the fields are given by
$N^2-1$ times the number of each fields, which are 
equivalent to $n_s=6$, $n_f=2$ and $n_v=1$. Then we find for large $N$,
\beq
 \langle T_{\mu}^{\mu}\rangle={N^2\over 32\pi^2}\left(
       R^{\mu\nu}R_{\mu\nu}-{1\over 3}R^2\right)=-{3\lambda^2\left(1+{\dot{\lambda}\over 2\lambda}{a_0\over \dot{a}_0}\right)\over 8\pi^2}N^2\, .
       \label{anomaly3}
\eeq
This result (\ref{anomaly3}) is precisely equivalent to the above holographic one (\ref{anomaly2}).
Thus we could see that the holographic analysis could give correct results for the energy
momentum tensor even if the metric is time dependent as shown previously in \cite{GN13}.

\def\theequation{B.\arabic{equation}}
\setcounter{equation}{0}

\noindent
\section{ $\langle T_{\mu\nu}^{\rm IR}\rangle$ of the dual theory at $r=0$} \label{sec:app}

In the IR side, we get  $\langle T_{\mu\nu}^{\rm IR}\rangle$ by the parallel method.
By using the above formula (\ref{Feff-5}),
we find 
\bea
 \langle T_{\mu\nu}^{\rm IR}\rangle &=&{4R^3\over 16\pi G_N^{(5)}}~
           \left(\hat{g}_{(0)00}~t_{00},~\hat{g}_{(0)ij}~t_{11}\right)\, \label{stressIR-1} \\
t_{00} &=&-{3r_0^8\over {r^*}^4+\tilde{c}_0}\, , \quad 
  t_{11}=-r_0^8{2{r^*}^4+({r^*}{r^*}_1)^2+\tilde{c}_0 \over ({r^*}^4+\tilde{c}_0)(({r^*}{r^*}_1)^2-\tilde{c}_0)
}\, , \label{stressIR-2}
\eea 
This result should be interpreted as the VEV of the energy momentum {tensor} of the SYM theory 
living in the space-time $\hat{g}_{(0)\mu\nu}$ given by (\ref{Feff2IR}). In the present case, 
however, both the metric
 $\hat{g}_{(0)\mu\nu}$ and the $\langle T_{\mu\nu}\rangle$ are different 
from the one given at the boundary $r\to\infty$. Then we must check how the two theories on the 
each boundary are different. We perform this for the following three cases.

\vspace{.5cm}
One expect that the central charges on each boundaries would be different from each other since 
the renormalization group flow would be different. The answer for this issue is given by observing
the trace anomaly, which is found from the above $\langle T_{\mu\nu}\rangle$ as follows,
\bea
 \langle T^{\mu}_{\mu}\rangle &=&{4R^3\over 16\pi G_N^{(5)}}~
           \left(t_{00}+3t_{11}\right)\, \label{stressIR-3} \\
         &=&-{N^2 \over 2\pi^2}{6r_0^8{r^*}^2(({r^*}^2+{r^*}_1)^2) \over ({r^*}^4+\tilde{c}_0)(({r^*}{r^*}_1)^2-\tilde{c}_0)
}\, , \label{stressIR-4}
\eea
where we used $G_N^{(5)}=8\pi^3{\alpha'}^4g_s/R^5$ and $R^4=4\pi N{\alpha'}^2g_s$. This is rewritten by using
the relation
\bea
   W_{IR}&=&\left( R^{\mu\nu}R_{\mu\nu}-{1\over 3}R^2\right)\,  \label{weyl-1} \\
    &=& -16 {6r_0^8 {r^*}^2({r^*}^2+{r^*}_1^2 ) \over ({r^*}^4+\tilde{c}_0)(({r^*}{r^*}_1)^2-\tilde{c}_0)}\, , \label{weyl-2}
\eea
we obtain
\beq
         \langle T^{\mu}_{\mu}\rangle={N^2\over 32\pi^2}W_{IR}\,  . \label{stressIR-5}
\eeq

\vspace{.3cm}
\section*{Acknowledgments}
This work of M. I was supported by World Premier International Research Center Initiative （WPI）, MEXT, Japan.

\vspace{1cm}

\newpage
\end{document}